# LIFE-DETECTION TECHNOLOGIES FOR THE NEXT TWO DECADES

Chaitanya Giri[1,2*], Tony Jia[1], H. James Cleaves II[1,3,4,5], Tomohiro Usui[1,6], Dhananjay Bodas[7], Christopher Carr[8,9], Huan Chen[10], Yuka Fujii[1], Yoshihiro Furukawa[11], Hidenori Genda[1], Richard J. Gillams[1], Keiko Hamano[1], Shingo Kameda[12], Ramanarayanan Krishnamurthy[13], Cornelia Meinert[14], Markus Meringer[15], Kishore Paknikar[7], Sudha Rajamani[16], Sisinthy Shivaji[17], Andrew Steele[2], Masateru Taniguchi[18], Hikaru Yabuta[19], Akihiko Yamagishi[20]

1. Earth-Life Science Institute, Tokyo Institute of Technology, 2-12-1-IE-1 Ookayama, Meguro-ku, Tokyo 152-8550, Japan
2. Geophysical Laboratory, Carnegie Institution for Science, Washington, DC 20015, USA
3. Blue Marble Space Institute of Science, Washington, DC 20011, USA
4. Center for Chemical Evolution, Georgia Institute of Technology, Atlanta, GA 30332, USA
5. Institute for Advanced Study, 1 Einstein Drive, Princeton, NJ 08540, USA
6. Department of Earth and Planetary Sciences, Tokyo Institute of Technology, 2-12-1 Ookayama, Meguro-ku, Tokyo 152-8551, Japan
7. Nanobioscience Group, Agharkar Research Institute, GG Agarkar Road, Pune 411004, India
8. Earth, Atmospheric and Planetary Science, Massachusetts Institute of Technology, 77 Massachusetts Ave, Cambridge, MA 02139, USA
9. Department of Molecular Biology, Massachusetts General Hospital, 185 Cambridge Street, Boston, MA 02114, USA
10. National High Magnetic Field Laboratory, Florida State University, 1800 East Paul Dirac Drive, Tallahassee, FL 32310, USA
11. Department of Earth Science, Tohoku University, 6-3 Aramaki Aza Aoba, Aoba-ku, Sendai 980-8578, Japan
12. Department of Physics, Rikkyo University, 3-34-1 Nishi-Ikebukuro, Toshima, Tokyo 171-8501, Japan
13. Department of Chemistry, The Scripps Research Institute, 10550 North Torrey Pines Road, La Jolla, CA 92037, USA
14. Université Côte d'Azur, CNRS, Institut de Chimie de Nice UMR 7272, Nice, France
15. German Aerospace Center (DLR), Earth Observation Center (EOC), Münchner Straße 20, 82234 Oberpfaffenhofen-Wessling, Germany.
16. Indian Institute for Science Education and Research (IISER) Pune, Maharashtra 411008, India
17. LV Prasad Eye Institute, LV Prasad Marg, Banjara Hills, Hyderabad, Andhra Pradesh 500034, India
18. The Institute of Scientific and Industrial Research, Osaka University, 8-1 Mihogaoka, Ibaraki, Osaka 567-0047, Japan
19. Department of Earth and Planetary Systems Science, Hiroshima University, 1-3-1 Kagamiyama, Hiroshima 739-8526, Japan
20. Department of Applied Life Sciences, Tokyo University of Pharmacy and Life Sciences, Hachioji, 192-0355 Tokyo, Japan

*Submitted to and Cited by the National Academy of Sciences in support of the Astrobiology Science Strategy for the Search for Life in the Universe*

# 1. LIFE-DETECTION: A CENTRAL RATIONALE FOR SPACE EXPLORATION

Since its inception six decades ago, astrobiology has diversified immensely to encompass several scientific questions including the origin and evolution of Terran life, the organic chemical composition of extraterrestrial objects, and the concept of habitability, among others. The detection of life beyond Earth forms the main goal of astrobiology, and a significant one for space exploration in general. This goal has galvanized and connected with other critical areas of investigation such as the analysis of meteorites and early Earth geological and biological systems, materials gathered by sample-return space missions, laboratory and computer simulations of extraterrestrial and early Earth environmental chemistry, astronomical remote sensing, and in-situ space exploration missions. Lately, scattered efforts are being undertaken towards the R&D of the novel and as-yet-space-unproven 'life-detection' technologies capable of obtaining unambiguous evidence of extraterrestrial life, even if it is significantly different from Terran life [1]. As the suite of space-proven payloads improves in breadth and sensitivity, this is an apt time to examine the progress and future of life-detection technologies.

# 2. ELSI-EON WORKSHOP ON LIFE DETECTION TECHNOLOGIES

The past four National Aeronautics and Space Administration (NASA) Astrobiology Roadmap documents acknowledged the need to develop technologies that can unambiguously detect life on habitable planetary bodies [2]. These roadmaps also mention the importance of assessing habitability and biosignature preservation potential, searching for liquid water, and defining thermodynamic constraints as critical parameters for selecting planetary targets and sites for future life-detection missions.

The Earth-Life Science Institute (ELSI) is a vital research center for trans-disciplinary scientists across the world working towards the grand scientific questions of understanding the formation of the Earth, the origins of life, and the evolution of inhabited and habitable objects in the solar system and elsewhere in the universe. An international workshop entitled "Life Detection Technology: For Mars, Enceladus, and Beyond" was organized on October 5-6, 2017 at ELSI; the co-authors of this white paper were the participants. The purpose of the workshop was to **(a)** deliberate the utilities of diverse life-detection payloads on space probes for exploring planetary bodies in the solar system with dissimilar habitability potential; **(b)** cultivate international synergies between scientific and engineering laboratories from around the world for efficient R&D of life-detection technologies; and **(c)** add to the transdisciplinarity of this domain by bringing in new scientific and engineering disciplines that are presently outside astrobiology and could assist in the development of life-detection technologies. This white paper summarizes the discussions that emerged from this workshop, which included participants from France, Germany, India, Japan and the United States. The participants presented their perspectives on what might constitute a signature of life, and what technologies might enable such detection.

Among the leading questions in astrobiology are: What is life? How do we define life? Will life elsewhere be identical or similar to Terran life? And what planetary environmental parameters determine habitability? These are presently studied from various physical, chemical, and biological perspectives [3]. Since these are grand scientific questions, they are difficult to tackle from the narrow purview of stove-piped scientific disciplines. Astrobiology and instrument-driven life detection stimulate transdisciplinarity, to which the congregation of this workshop was testament. The emergence of technical capacities to explore the surfaces of planetary bodies through space probes, e.g., landers, rovers and orbiters, are revitalizing the possibility of answering these questions. The space agencies in the United States, Japan, India, and the European Union pursue the R&D of space payloads dedicated to the search of bio-geo-chemo-signatures. However, presently none of these payloads are capable of detecting life. As life-detection technologies become increasingly central for in-situ explorations, they will significantly advance our scientific understanding of the possibilities of life to survive beyond Earth, even simultaneously on multiple celestial objects. Therefore, apart from the basic scientific research, it is also essential to contemplate the technical demands of life-detection. The deliberations from this workshop are succinctly presented in the following sections.

## 3. HABITABILITY ON PLANETARY AND NICHE SCALES

The habitability of any planet or satellite is estimated from its size, surface composition, climate, orbit, and exposure to stellar radiation, among other parameters. Prior knowledge of the events that each planetary body experiences during its formation is also imperative. For example, events such as bolide impact-driven degassing of volatiles from planetary interiors may result in rapid retention of liquid water on the planetary surface and a gaseous atmosphere. With the advent of next-generation astronomical observatories like the James Webb Space Telescope, the Extremely Large Telescope, and the Wide Field Infrared Survey Telescope, the theoretical knowledge of planetary habitability will receive support from a sizeable statistical set of spectrally-characterized extrasolar planets. A reasonably-sized sample set of extrasolar planets will potentially contribute to our understanding of the possibility (or possible forms) of life existing on them. The telescopes and other instruments used for the characterization of habitable extrasolar planets will also characterize habitable bodies in the solar system at much higher spatial resolutions. These intra-solar system observations will be necessary for selecting landing site, a crucial factor shaping the type of life-detection payloads aboard exploration probes.

Lessons from the traditional laboratory-based prebiotic chemistry research can tentatively inform the search for potential extraterrestrial biosignatures, but reliance on these studies to determine reasonable biomarkers must be considered critically. Extant biochemistry is presumably a product of the molecular evolution of various chemical species that might have populated the prebiotic soup. This process was perhaps driven by pertinent selection pressures across millennia, which eventually resulted in life as we know it today. Present prebiotic chemistry research is heavily biased by our knowledge of extant biochemistry. It is vital to acknowledge that "acceptable" biosignatures could be biopolymers that have never been achieved in prebiotic chemistry research, not to mention that chemistries in the sterile, controlled laboratory may be very different from chemistries in the field. A complete bias towards finding extant Terran biosignatures when searching for life elsewhere should, therefore, be avoided.

It is supposed that the most convincing biosignatures are likely to be organic, simply because carbon is uniquely able to form a vast structural and informational molecular repertoire. Life-detection techniques targeting a wide array of carbon-based molecules can be applied to all samples, those existing in the same or different environments or even environments undergoing temporal variation. The insights obtained from such an approach is that extant biological, abiological or extinct biological samples will all provide unique identifiable signals. Biological and abiological samples may both contain thousands or millions of unique low molecular mass chemical species, and these can be explored in depth in the laboratory. Even if the identities of these species are not entirely known, the relationships between them can be indicative of biology. This aspect could be especially useful for extraterrestrial life-detection, as it is possible the nature of terrestrial biochemistry is either historically contingent or tightly linked to Earth's geochemistry, and thus alien life could have evolved differently. Terran life produces a unique ensemble of organic molecules that is distinct from the vast combinatorial chemical space of abiotic chemistry. To maximize the chances of identifying real biosignatures and avoid false positives, an approach targeting chemical distributions to identify patterns unique to life will be necessary.

## 4. CONCEPTS & TECHNOLOGIES UNDER CONSIDERATION

Life-detection demands a technologically intricate space mission design. One causal factor of this intricacy is the fact that habitable environments are not distributed globally on planetary bodies, but possibly exist in geographically limiting niches. Reaching such often-inaccessible sites will require agile robotic probes that are robust, able to seamlessly communicate with orbiters and deep space communications networks, be operationally semi-autonomous, have high-performance energy supplies, and are sterilizable to avoid forward contamination. Moreover, to build confidence in any positive detection of life beyond Earth, cutting-edge payloads are needed that can investigate multiple aspects of the 'Life Detection Ladder' described previously [4].

Despite their potential habitability, the environmental conditions on Enceladus, Mars, and other planetary bodies are dissimilar to Earth and hence pose challenges for the R&D of appropriate life-detection instruments. Even assuming that life could exist in all these places, the workshop participants noted that a probe-payload combination designed for a mission to a potentially habitable niche on one planetary body would not work seamlessly for niches on another body. Given the distinct biology or bio-chemo-markers that different environments sustain, thus the probe-payload combination and the space mission design needed to explore habitable zones on Mars, Enceladus, Titan, and Europa would need to be custom-made.

In agreement with the suggestions of the NASA Life Detection Ladder, the participants in this workshop promoted a variety of life-detection instruments. In-situ visual recognition of micro-organisms and detection of genetic or metabolic bio-macromolecules are some of the current aims of extant life detection technologies. The bio-geo-chemo-signatures of extinct and extant life can be detected using Raman and other spectroscopy techniques, enantioselective and two-dimensional gas chromatography, high-resolution mass spectrometry, microfluidic devices, and microscopes. The workshop participants agreed on the necessity to pursue life-detection space missions with a suite of several instruments. Results obtained from various instruments can avoid spurious measurements and provide statistical analysis.

To search for life in regions theoretically devoid of life requires novel detection techniques or probes. For example, air sampling in Earth's stratosphere with a novel scientific cryogenic payload has led to the isolation and identification of several new species of bacteria; this was an innovative technique analyzing a region of the atmosphere that was initially believed to be devoid of life [5]. Novel high-sensitivity fluorescence microscopy techniques may be utilized to detect extraterrestrial organic compounds with catalytic activity surrounded by membranes, i.e., extraterrestrial cells [6]. Nucleic acid (i.e., genetic/informational biopolymers) detection and sequencing [7] provides an even more unambiguous approach to detecting ancestrally related life, Terran contamination, or non-familiar nucleic acid-based life. Despite the advent of highly portable single-molecule sequencing technology, current methods require extensive conditioning of nucleic acid molecules (library preparation) and biological reagents. Technologies under development, such as quantum tunneling-based nanogap devices [8], could eliminate this complexity and simultaneously target nucleic acids, peptides, and other small molecules while achieving improved detection limits and broadening the potential range of life that could be detected. Incorporating microfluidics—due to their requirement of small fluid volumes, miniaturization, and low power consumption—that use novel nanomaterials for identifying microorganisms or their signature molecules is an ideal proposition for space missions which have weight and size constraints. Enantioselective separation techniques can distinguish between amino acids and sugars formed by abiotic or biotic reaction mechanisms and detect molecular homochirality, which may be a diagnostic biosignature [9]. Enantioselective gas chromatography has been utilized on the ESA Rosetta and ExoMars and NASA Mars Science Laboratory missions. It can be used with pertinent innovation for future life-detection missions.

Mass spectrometry (MS) has been extensively used for surface and atmospheric chemical characterization on numerous space missions. Miniaturized mass spectrometers with increased mass resolution and multiple steps of fragmentation (e.g., Cosmo-Orbitrap by European Space Agency (ESA), MULTUM by Japan Aerospace Exploration Agency (JAXA), and MASPEX and LD-TOF-MS by NASA) will be available for in-situ measurements on future life-detection missions. These MS techniques would allow characterization of high-mass organic solids including biopolymers and also enable in-situ elemental composition measurement for mineralogy and isotopic dating methods [10], which are essential for characterizing geo-chemo-signatures of habitability. The exciting developments in machine learning and its application to complex MS data will be invaluable in aiding the detection of organic and inorganic markers of biology. The specificity of these and other instruments also suggest life-detection missions demands the continuous invention of novel probe-payload combinations customized for exploration of each potentially habitable site. In the 2020s, sample-return missions like JAXA's Martian Moons eXploration mission to Phobos and Hayabusa-2 to asteroid Ryugu and NASA's OSIRIS-rEX to asteroid Bennu will refurbish the Earth-based infrastructure for environmentally-controlled and near-sterile curation and analyses of organic-enriched extraterrestrial materials. The sample handling knowledge generated from these missions will improve planetary protection procedures. Along with the advances anticipated from in-situ exploration, sample-return missions will also contribute to advances in handling potentially biotic extraterrestrial materials.

Sample-return missions are inherently technically sophisticated, but high-performance ground-based instruments can extensively characterize returned samples. High-resolution analyses on in-situ exploration missions are presently challenging from the purview of data transmission rate, as huge amounts of data may be generated. These aspects of life-detection missions call for the advancement of the current deep space communication technologies.

Analytical instruments associated with high-powered synchrotron radiation and magnetic field facilities will continue to possess superior characterization abilities, and only through sample-return missions, their features could be utilized. Ultra-high resolution Fourier-transform-ion cyclotron resonance-MS supported by high magnetic fields allows unambiguous assignment of molecular formulas to samples containing high molecular mass organic solids and polymers. Another technique, the synchrotron-based scanning transmission x-ray absorption microscopy is capable of distinguishing the distributions of protein, polysaccharide, and lipid in a living microorganism, and also characterizing biomineralization and nano-scale bioweathering. These techniques are disposed to provide more reliable and comprehensive characterization of chemically-complex materials. Efforts are also being undertaken to process high-resolution chemical characterization data with pattern recognition, machine learning, and artificial intelligence to determine the biological or abiological origin of the samples, a crucial determinant of the presence of life.

## 5. CONCLUSION

The authors of this white paper unanimously recognize the significance of life-detection instruments for unambiguous identification of extraterrestrial life and addressing the challenges involved in this. The authors acknowledge the necessity to establish an international network to forge collaborative R&D of life-detection technologies and a worldwide peer-reviewing network for data analyses. Life-detection is a capital-intensive endeavor capable of yielding enormous scientific return-on-investment and industrial spin-offs. An international network is crucial for pooling and coordinating human, financial, and technical resources and harnessing creativity, talent, and infrastructure across institutions and governments. These factors will be vital for the R&D of life detection technologies and the growth of astrobiology as a science in the decades to come.

## ACKNOWLEDGEMENTS

The authors are grateful to the Earth-Life Science Institute (ELSI) for supporting the workshop on which this white paper is based. The ELSI, at the Tokyo Institute of Technology, is a World Premier International Research Center supported by the Ministry of Education, Culture, Sports, Science and Technology (MEXT) of the Government of Japan. The workshop and the publication are supported by ELSI and the ELSI Origins Network (EON), which is supported by a grant from the John Templeton Foundation. The opinions communicated in this white paper are those of the authors and do not essentially reflect the views of the John Templeton Foundation.